\documentclass[twocolumn]{aa}
\usepackage{graphicx,epsfig}

\newcommand{\fgas}{f_{\rm gas}}

\newcommand{\Ho}{H_{\rm o}}
\newcommand{\OmM}{\Omega_{\rm M}}
\newcommand{\OmL}{\Omega_\Lambda}
\newcommand{\OmB}{\Omega_{\rm B}}

\newcommand{\thetac}{\theta_{\rm c}}
\newcommand{\thetav}{\theta_{\rm v}}
\newcommand{\thetaf}{\theta_{\rm f}}

\newcommand{\Yo}{Y_{\rm o}}

\newcommand{\fwhm}{\theta_{\rm fwhm}}

\newcommand{\yo}{y_{\rm o}}	
\newcommand{\jnu}{j_\nu}		
\newcommand{\nui}{\nu_i}
\newcommand{\Psit}{\Psi_{\thetac}}
\newcommand{\Tc}{T_{\thetac}}
\newcommand{\sigt}{\sigma_{\thetac}}
\newcommand{\Ft}{F_{\thetac}}
\newcommand{\Pnoise}{P^{\rm noise}}
\newcommand{\Psky}{P^{\rm sky}}

\begin{document}
   \title{Catalog Extraction in SZ Cluster Surveys: 
		a matched filter approach}

   \subtitle{}

   \author{Jean--Baptiste Melin\inst{1,2}
           \thanks{\emph{New address: } CEA Saclay, DAPNIA/SPP, 91191 Gif-sur-Yvette, FRANCE}
           \and James G. Bartlett\inst{1}
	  \and Jacques Delabrouille\inst{1}
          }

   \offprints{J.--B. Melin}

   \institute{\inst{1}APC, 11 pl. Marcelin Berthelot, 75231 
                     Paris Cedex 05, FRANCE \\
			(UMR 7164 -- CNRS, Universit\'e
                    Paris 7, CEA, Observatoire de Paris)\\
              \inst{2}Department of Physics, University of California Davis,
                     One Shields Avenue, Davis, CA , 95616 USA\\
             \email{jean-baptiste.melin@cea.fr, bartlett@apc.univ-paris7.fr, delabrouille@apc.univ-paris7.fr}
             }

   \date{}

  \abstract{We present a method based on matched multifrequency filters
for extracting cluster catalogs from Sunyaev--Zel'dovich (SZ) surveys.  We evaluate
its performance in terms of completeness, contamination rate and photometric
recovery for three representative types of SZ survey: a high resolution 
single frequency radio survey (AMI), a high resolution ground--based 
multiband survey
(SPT), and the Planck all--sky survey.  These surveys are not purely flux limited,
and they loose completeness significantly before their point--source detection
thresholds.  Contamination remains relatively low at $<$~5\% (less than 30\%)
for a detection threshold set at S/N=5 (S/N=3).  We identify photometric recovery
as an important source of catalog uncertainty: dispersion in recovered flux from
multiband surveys is larger than the intrinsic scatter in the $Y-M$ relation 
predicted from hydrodynamical simulations, while photometry in the single frequency
survey is seriously compromised by confusion with primary cosmic microwave 
background anisotropy.  The latter effect implies that follow--up observations in other 
wavebands (e.g., 90~GHz, X--ray) of single frequency
surveys will be required.  Cluster morphology can cause a bias in 
the recovered $Y-M$ relation, but has little effect on the scatter;
the bias would be removed during calibration of the relation.
Point source confusion only slightly decreases multiband survey 
completeness; single frequency survey completeness could be significantly reduced
by radio point source confusion, but this remains highly uncertain because we
do not know the radio counts at the relevant flux levels.
  }
\titlerunning{Catalog Extraction in SZ Cluster Surveys}
\authorrunning{Melin et al.}
   \maketitle
%

\section{Introduction}
Galaxy cluster catalogs play an important role in cosmology
by furnishing unique information on the matter distribution
and its evolution.  Cluster catalogs, for example, efficiently 
trace large--scale features, such as the recently detected 
baryon oscillations (\cite{eisenstein_etal05,
cole_etal05,angulo_etal05,huetsi05}), and provide a
sensitive gauge of structure growth back to high redshifts 
(\cite{ob92,rosati_etal02,voit04} and references therein). 
This motivates a number of ambitious projects proposing to use
large, deep catalogs to constrain both galaxy evolution models and the 
cosmological parameters, most notably the dark energy abundance 
and equation--of--state (\cite{haiman_etal00,well_bat03,wang_etal04}).  
Among the most promising are 
surveys based on the Sunyaev--Zel'dovich (SZ) effect (\cite{sun70,sun72} and
see \cite{bir99,car02} for reviews), 
because it does not suffer from surface brightness dimming and because 
we expect the observed SZ signal to tightly correlate to cluster 
mass (\cite{bar01,motl_etal05}). Many authors have investigated the scientific 
potential of SZ surveys to constrain cosmology (e.g., \cite{barb96,col97,hol00,
kne01,ben02}), emphasizing the advantages intrinsic to observing 
the SZ signal.  

Cosmological studies demand statistically pure catalogs with well 
understood selection criteria.  As just said, SZ surveys are intrinsically
good in this light; however, many other factors -- related, for example,
to instrumental properties, observing conditions, astrophysical 
foregrounds and data reduction algorithms -- influence 
the selection criteria.  This has prompted some authors 
to begin more careful scrutiny of SZ survey selection functions in 
anticipation of future observations
(\cite{bar01,sch03,white03}, Vale \& White~\cite{vale05}, Melin et al.~\cite{mel05}, \cite{juin05}).  

In Melin et al.~(\cite{mel05}), we presented a general formalism for the 
SZ selection function together with some preliminary applications
using a matched--filter cluster detection method.
In this {\em paper} we give a thorough presentation of
our cluster detection method and evaluate its performance 
in terms of catalog completeness, contamination and photometric recovery.
We focus on three types of SZ survey: single frequency radio surveys like
the Arcminute MicroKelvin Imager (AMI interferometer) survey\footnote{\tt http://www.mrao.cam.ac.uk/telescopes/ami/}, multi--band ground--based bolometric surveys such as
the South Pole Telescope (SPT) survey\footnote{\tt http://astro.uchicago.edu/spt/}, and the space--based Planck survey\footnote{\tt http://astro.estec.esa.nl/Planck/}. 
In each case, we quantify the selection function using the formalism of 
Melin et al.~(\cite{mel05}).

We draw particular attention to the oft--neglected issue of photometry.  
Even if the SZ flux--mass relation is intrinsically tight, what matters in 
practice is the relation between the {\em observed} SZ flux and
the mass.  Photometric errors introduce both bias and additional 
scatter in the observed relation.  Calibration of the $Y-M$ relation
will in principal remove the bias; calibration precision, however, 
depends crucially on the scatter in the observed relation.  
Good photometry is therefore very important.  As we will see, 
observational uncertainty dominates the predicted intrinsic scatter in 
this relation in all cases studied.

We proceed as follows.
In section 2, we discuss cluster detection techniques and present 
the matched filter formalism.  We describe our detection
algorithm in Section 3. Using Monte Carlo simulations of the three
types of survey, we discuss catalog completeness, contamination and photometry. This is done in Section 4 under the ideal situation
where the filter perfectly matches the simulated clusters and in 
the absence of point sources.  In Section 5 we examine effects 
caused by cluster morphology, using N--body simulations, and then 
the effect of point sources.  
We close with a final discussion and conclusions in Section 6.

\section{Detecting Clusters}

The detection and photometry of extended sources presents a 
complexity well appreciated in Astronomy.  Many powerful 
algorithms, such as {\it SExtractor} (\cite{ber96}), have 
been developed to extract extended sources superimposed on
an unwanted background.  They typically estimate the local background
level and group pixels brighter than this level into
individual objects.  Searching for clusters at millimeter wavelengths
poses a particular challenge to this approach, because the clusters
are embedded in the highly variable background of the primary CMB 
anisotropies and Galactic emission. 
Realizing the importance of this issue, several authors have proposed 
specialized techniques for SZ cluster detection.  Before detailing our
own method, we first briefly summarize some of this work in order to
motivate our own approach and place it in context.

\subsection{Existing Algorithms}

Diego et al.~(\cite{die02})
developed a method designed for the Planck mission that is based on
application of {\em SExtractor} to SZ signal maps constructed by combining
different frequency channels.  It makes no assumption about
the frequency dependance of the different astrophysical signals, nor
the cluster SZ emission profile.  The method, however, requires many 
low--noise maps over a broad range of frequencies in order to construct
the SZ map to be processed by {\it SExtractor}.  Although they will
benefit from higher resolution, planned ground--based surveys will 
have fewer frequencies and higher noise levels, making application of 
this method difficult.

In another approach, Herranz et al. (\cite{her02a}, \cite{her02b}; see also
\cite{lop05} for point--source applications) 
developed an ingenious filter ({\it Scale Adaptive Filter}) 
that simultaneously extracts cluster size and flux.
Defined as the optimal filter for a map containing a single cluster, it
does not account for source blending.  Cluster--cluster blending 
could be an important source of confusion in 
future ground--based experiments, with as a consequence poorly estimated
source size and flux.  

Hobson \& McLachlan~(\cite{hob03}) recently 
proposed a powerful Bayesian detection method using a Monte Carlo 
Markov Chain. The method simultaneously solves for the position, size, 
flux and morphology of clusters in a given map.  Its 
complexity and run--time, however, rapidly increase with the number 
of sources.  

More recently, Sch{\"a}fer et al.~(\cite{sch04}) generalized scale adaptive and
matched filters to the sphere for the Planck all--sky SZ survey.
Pierpaoli et al.~(\cite{pier04}) propose a method based on wavelet filtering,
studying clusters with complex shapes.
Vale \& White~(\cite{vale05}) examine
cluster detection using different filters (matched, wavelets, mexican
hat), comparing completeness and contamination levels.

Finally, Pires et al.~(\cite{pires05}) 
introduced an independent component analysis 
on simulated multi--band data to separate the SZ signal, followed by
non--linear wavelet filtering and application of SExtractor.  

\begin{figure*}
\begin{center}
\includegraphics[scale=1]{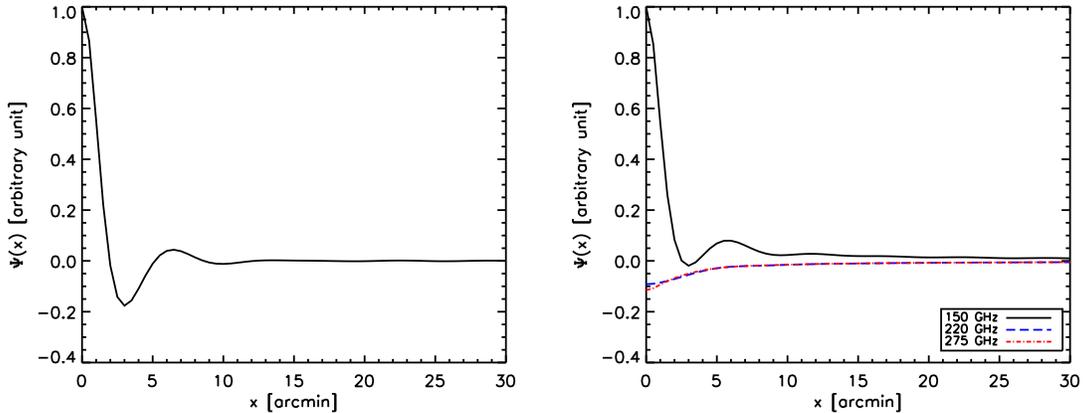}
\caption{Two examples of the matched filter for $\thetac=1$~arcmin.  
The curves give the radial
profiles of the filters, which are symmetric because we have chosen
a symmetric cluster template.  {\em Left:} 
filter for a single frequency survey with a $\fwhm=1.5$~arcmin beam 
and 8~$\mu$K instrumental noise/beam (AMI--like, see Table~\ref{tab:experiments}). The undulating form of the filter maximizes the cluster signal while reducing contamination from primary CMB anisotropy. 
{\em Right:}  The three components of the 
3--band filter for a SPT--like experiment (Table~\ref{tab:experiments}).
The filter is arbitrarily normalized to unity at 150~GHz.
The filter uses both spatial and frequency weighting to 
optimally extract the cluster signal from the CMB and instrument noise.
Although in this Figure the filters continue to large radii, in practice we
truncate them at $10\thetac$.
}
\label{fig:filters}
\end{center}
\end{figure*}

Our aim is here is two--fold: to present and extensively evaluate 
our own SZ cluster catalog extraction method, and to use it in a 
comprehensive study of SZ survey selection effects.  The two are in
fact inseparable.  First of all, selection effects are specific to a 
particular catalog extraction method.  Secondly, we require a 
robust, rapid algorithm that we can run over a large number of simulated
data sets in order to accurately quantify 
the selection effets.  This important
consideration conditions the kind of extraction algorithm that we can
use.  With this in mind, we have developed a fast catalog construction
algorithm based on matched filters for both single and multiple frequency
surveys.  It is based on the approach first proposed by Herranz et al.,
but accounts for source blending.

After describing the method, we apply the formalism given in 
Melin et al.~(\cite{mel05}) to quantify the selection function and 
contamination level in up--coming SZ surveys.
We take as representative survey configurations AMI, SPT and
Planck, and Monte Carlo simulate the entire catalog
extraction process from a large ensemble of realizations for each 
configuration.  
By comparing to the simulated input catalogs, we evaluate 
the extracted catalogs in terms of their completeness, 
contamination and photometric accuracy/precision.  We will place
particular emphasis on the importance of the latter, something 
which has received little attention in most studies of this kind.

\subsection{Matched Filters}

The SZ effect is caused by the hot gas ($T\sim 1-10$~keV) contained in
galaxy clusters known as the intracluster medium (ICM); electrons in
this gas up--scatter CMB photons and create a unique spectral distortion
that is negative at radio wavelengths and positive in the
submillimeter, with a zero--crossing near 220~GHz.  The form of this
distortion is universal (in the non--relativistic limit applicable to
most clusters), while the amplitude is given by the Compton $y$
parameter, an integral of the gas pressure along the line--of--sight.
In a SZ survey, clusters will appear as sources extended over
arcminute scales (apart from the very nearby objects, which are
already known) with brightness profile
\begin{equation}
\Delta i_\nu(\vec{x}) = y(\vec{x}) \jnu
\end{equation}
relative to the mean CMB brightness.
Here $y(\vec{x})$ is the Compton $y$ parameter at position $\vec{x}$ (a
2D vector on the sky) and $\jnu$ is the SZ spectral function evaluated at 
the observation frequency $\nu$.

Matched filters for SZ observations were first proposed by
Haehnelt \& Tegmark~(\cite{hae96}) as a tool to estimate cluster peculiar
velocities from the kinetic effect, and Herranz et al. (\cite{her02a},
\cite{her02b}) later showed how to use them to detect
clusters via the thermal SZ effect.  They are designed  
to maximally enhance the signal--to--noise for a SZ cluster source
by optimally (in the least square sense) filtering the data, which in our case is a sky map or 
set of maps at different frequencies.  They do so by incorporating 
prior knowledge of the cluster signal, such as its spatial
and spectral characteristics.  The unique and universal frequency 
spectrum of the thermal SZ effect (in the non--relativistic regime)
is hence well suited for a matched--filter approach.  

\begin{figure}[htb]
\includegraphics[scale=0.5]{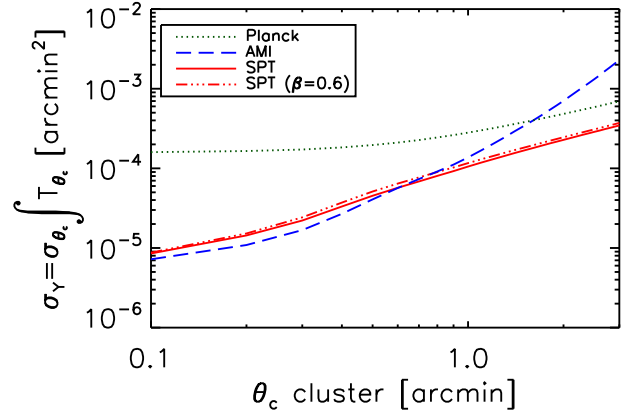}
\caption{Filter noise expressed in terms of integrated SZ flux $Y$ -- 
$\sigma_{\rm Y} = \sigma_{\theta_c} \int T_{\theta_c}({\vec x}) \, d{\vec x}$ -- as
a function of template core radius $\thetac$ for the three experiments 
listed in Table~\ref{tab:experiments}.  A cluster with $Y=\sigma_{\rm Y}$ 
would be detected at a signal--to--noise ratio $q=1$.  At a fixed 
detection threshold $q$ (e.g., 3 or 5), the completeness of a survey
rapidly increases from zero to unity in the region above its 
corresponding curve $q\sigma_{\rm Y}(\thetac)$ (Melin et al.~\cite{mel05}).
All the curves adopt our fiducial value of $\beta=2/3$, except 
the dashed--triple--dotted red curve, shown for comparison, which corresponds to the SPT case with $\beta=0.6$; 
this curve is systematically higher by ($2.5$ to $13)\%$, depending on $\theta_c$.
}
\label{fig:cy_tc}
\end{figure}

Less clear is the choice of the spatial profile $\Tc(\vec{x})$
to adopt for cluster SZ emission.  
One aims to choose a spatial 
template that represents as well as possible the average SZ emission 
profile. In other words, we want 
$\Tc(\vec{x})=\langle y(\vec{x})/\yo\rangle_C$, where the
average is over many clusters of size $\thetac$.
In the following, we
choose to describe clusters with a projected spherical $\beta$--profile:
\begin{equation}
y(\vec{x}) = \yo (1+|\vec{x}|^2/\thetac^2)^{-(3\beta-1)/2}
\end{equation}
with $\beta=2/3$ (with one exception, shown for comparison
in Figure~\ref{fig:cy_tc}).  
The spatial template is therefore 
described by a single parameter, the core radius $\thetac$; in 
our calculations, we truncate the profile at $10\thetac$.
This is a reasonable  choice, given X--ray observations 
(\cite{arnaud05}) of the intracluster medium and the resolution of planned SZ surveys.

In reality, of course, we know neither this average profile precisely nor
the dispersion of individual clusters around it beforehand.
This is an important point, because our choice for the template will 
affect both the detection efficiency and photometric accuracy.  
Detection efficiency will be reduced if the template does not well
represent the average profile and, as will become clear below, the
photometry will be biased.  In general, {\em the survey 
selection function unavoidably suffers from uncertainty induced by 
unknown source astrophysics} (in addition to other sources of uncertainty). 

In the following, we first study (Section 4) the ideal case where the filters 
perfectly match the cluster profiles, i.e., we use the $\beta$--model
for both our simulations and as the detection template.  In a later
section (5), we examine the effects caused by non--trivial cluster
morphology, as well as by point source confusion.
 
Consider a cluster with core radius $\thetac$ and central
$y$--value $\yo$ positioned at an arbitrary point $\vec{x}_{\rm o}$ 
on the sky.  For generality, suppose that the
region is covered by several maps $M_i(\vec{x})$ at $N$ different 
frequencies $\nui$ ($i=1,...,N$).  We arrange the survey maps
into a column vector $\vec{M}(\vec{x})$ whose $i^{th}$ component is the
the map at frequency $\nui$; this vector reduces to a scalar map in
the case of a single frequency survey.  Our maps contain the 
cluster SZ signal plus noise:
\begin{equation}
\vec{M}(\vec{x}) = \yo\vec{\jnu}\Tc(\vec{x}-\vec{x}_{\rm o}) + 
	\vec{N}(\vec{x})
\end{equation}
where $\vec{N}$ is the noise vector (whose components are noise
maps at the different observation frequencies) and $\vec{\jnu}$ 
is a vector with components given by the SZ spectral function $\jnu$ 
evaluated at each frequency.  Noise in this context 
refers to both instrumental noise as well as 
all signals other than the cluster thermal SZ effect; it thus also
comprises astrophysical foregrounds, for example, the primary
CMB anisotropy, diffuse Galactic emission and extragalactic point sources.  

We now build a filter $\vec{\Psit}(\vec{x})$ (in general, a column
vector in frequency space) that returns an estimate, $\hat{\yo}$, of
$\yo$ when centered on the cluster:
\begin{equation}\label{eq:filter}
\hat{\yo} = \int d^2x\; \vec{\Psit}^t(\vec{x}-\vec{x}_{\rm o}) 
      \cdot \vec{M}(\vec{x})
\end{equation}
where superscript $t$ indicates a transpose (with complex conjugation
when necessary).  This is just a linear combination of the maps, each
convolved with its frequency--specific filter $(\Psit)_i$.
We require an unbiased estimate of the central $y$ value, so that
$\langle \hat{\yo} \rangle = \yo$, where the average here is over both 
total noise and cluster (of core radius $\thetac$) ensembles.  Building
the filter with the known SZ spectral form and adopted spatial template
optimizes the signal--to--noise of the estimate; in other words, the
filter is {\em matched} to the prior information.  The filter is now
uniquely specified by demanding a minimum variance estimate.
The result expressed in Fourier space (the flat sky approximation is
reasonable on cluster angular scales) is (Haehnelt \& Tegmark~\cite{hae96}, Herranz et al.~\cite{her02a}, Melin et al.~\cite{mel05}):
\begin{equation}
\vec{\Psit}(\vec{k}) = \sigt^2 \vec{P}^{-1}(\vec{k})\cdot \vec{\Ft}(\vec{k})
\end{equation}
where 
\begin{eqnarray}
\vec{\Ft}(\vec{k})   & \equiv & \vec{\jnu} \Tc(\vec{k})\\
\label{eq:sigt}
\sigt          & \equiv & \left[\int d^2k\; 
     \vec{\Ft}^t(\vec{k})\cdot \vec{P}^{-1} \cdot
     \vec{\Ft}(\vec{k})\right]^{-1/2}
\end{eqnarray}
with $\vec{P}(\vec{k})$ being the noise power spectrum, a matrix in
frequency space with components $P_{ij}$ defined by $\langle
N_i(\vec{k})N_j^*(\vec{k}')\rangle_N=P_{ij}(\vec{k})
\delta(\vec{k}-\vec{k}')$.    
The quantity $\sigt$ gives the total noise variance through the filter. 
When we speak of the signal--to--noise of a detection, we refer to
$\hat{\yo}/\sigt$. 

We write the noise power spectrum as a sum 
$P_{ij}=\Pnoise_i\delta_{ij}+B_i(\vec{k})B^*_j(\vec{k})\Psky_{ij}$, where
$\Pnoise_i$ represents the instrumental noise power in band $i$,
$B(\vec{k})$ the observational beam and $P^{\rm sky}_{ij}$ 
gives the foreground power (non--SZ signal) between channels $i$ and $j$.  
As explicitly written, we assume uncorrelated instrumental noise between observation 
frequencies.  Note that we treat the astrophysical foregrounds as 
isotropic, stationary random fields with zero mean.  The zero 
mode is, in any case, removed from each of the maps, and the model
certainly applies to the primary CMB anisotropy.  It should 
also be a reasonable model for fluctuations of other foregrounds 
about their mean, at least over cluster scales\footnote{We make no assumption about the Gaussianity of the fields; the estimator remains unbiased even if
they are not Gaussian, although optimality must be redefined in this case.}.  

Two examples of the matched filter for $\thetac=1$~arcmin 
are shown in Fig.~\ref{fig:filters}, 
one for an AMI--like single frequency survey with a 1.5~arcmin beam 
(left--hand panel) and the other for a SPT--like 3--band filter (right--hand
panel); see Table~\ref{tab:experiments} for the experimental characteristics.
The filters are circularly symmetric, with the figures giving their radial
profiles, because we have chosen a spherical cluster model.
We clearly see the spatial weighting used by the single frequency filter 
to optimally extract the cluster from the noise and CMB backgrounds.  
The multiple frequency filter $\vec{\Psit}$ is a 3--element column 
vector containing filters for each individual frequency.   
In this case, the filter employs both spectral and spatial
weighting to optimally extract the cluster signal.  

Figure~\ref{fig:cy_tc} shows the filter noise as a function of 
template core radius $\thetac$.  We plot the filter noise expressed in terms
of an equivalent noise $\sigma_{\rm Y}\equiv \sigma_{\theta_c} \int T_{\theta_c}({\vec x}) \, d{\vec x}$ on the integrated SZ flux $Y$. The dashed--triple--dotted red curve with $\beta=0.6$ is shown 
for comparison to gauge the impact of changing this parameter, otherwise fixed at $\beta=2/3$ 
throughout this work. Melin et al.~(\cite{mel05}) use
the information in this figure to construct survey completeness functions.  At fixed
signal--to--noise $q$, the completeness of a survey rapidly increases to
unity in the region above the curve $q\sigma_{\rm Y}$.  
The Figure shows that high angular resolution ground--based surveys 
(e.g., AMI, SPT) are not purely flux limited, because their noise
level rises significantly with core radius.  The lower resolution of the 
Planck survey, on the other hand, results in more nearly flux limited 
sample.  

\section{Catalog Extraction}
\label{sec:det}

Catalog construction proceeds in three steps, the last two of which
are repeated\footnote{Note that we have made some changes in the two last 
steps compared to the description given in~Melin et al.~(\cite{mel05}). We no longer 
sort candidates in a tree structure for de-blending; instead, we identify 
and then remove candidates one by one from the filtered maps.
This has only a small impact on the completeness of the detection 
algorithm, leaving the conclusions of our previous paper intact. The changes,
however, greatly improve photometry and lower contamination.}: 
\begin{enumerate}
\item Convolution of the frequency map(s) with matched filters 
corresponding to different cluster sizes;
\item Identification of candidate clusters as objects with signal--to--noise 
$\hat{\yo}/\sigt>q$, where $q$ is our fixed detection threshold, followed by photometry of the brightest remaining cluster candidate, which is then 
added to the final cluster catalog;
\item Removal of this object from the set of filtered maps using 
the photometric parameters (e.g., $\yo$ and $\thetac$) from the  
previous step.
\end{enumerate}
We loop over the last two steps until there are no remaining candidates
above the detection threshold.  The following sections detail each step.

\subsection{Map filtering}

In the first step, we convolve the observed map(s) with matched filters
covering the expected range of core radii.  For AMI and SPT, for example,
we vary $\theta_c$ from $0.1$ to $3 \, {\rm arcmins}$ in $0.1$ steps 
(i.e., $\theta_c = 0.1, 0.2,...,2.9, 3 \, {\rm arcmins}$) and add 
three values for the largest clusters: $4, 5, 6 \, {\rm arcmins}$. 
We thus filter the map(s) $n_{\theta_c}$ times ($n_{\theta_c}=33$ 
for AMI and SPT) to obtain $2 \, n_{\theta_c}$ filtered maps, 
$J_{\theta_c}$ et $L_{\theta_c}$.
The $n_{\theta_c}$ maps $J_{\theta_c}$ give the SZ amplitude
(obtained using $\Psit$), while the $n_{\theta_c}$  
maps $L_{\theta_c}$ give the signal--to--noise ratio: 
$L_{\theta_c}=J_{\theta_c}/\sigt$).  
We set a detection threshold at fixed signal--to--noise $q$
and identify candidates at each filter scale $\theta_c$ as pixels with 
$L_{\theta_c}>q$.  Common values for the threshold are $q=3$ and $q=5$; 
the choice is a tradeoff between detection
and contamination rates (see below).

\subsection{Cluster parameter estimation: Photometry}

We begin the second step by looking for the brightest candidate 
pixel in the set of maps $L_{\theta_c}$.  
The candidate cluster is assigned the spatial coordinates 
$(x,y)$ of this pixel, and its core radius is defined as the 
filter scale of the map containing the pixel: $\thetac=\thetaf$.  We then calculate the total integrated flux using 
$Y=\hat{\yo} \int T_{\theta_c}({\vec x}) \, d{\vec x}$, where $\hat{\yo}$ 
is taken from the map $J_{\rm \theta_c}$ at the same filter scale.  
We refer to this step as the photometric step, and the parameters $\hat{\yo}$, 
$\thetac$ and $Y$ as photometric parameters.  Note that measurement error
on $Y$ comes from errors on both $\hat{\yo}$ and $\thetac$ (We 
return to this in greater detail in Section~\ref{sec:photo}).

\subsection{Catalog construction}

The candidate cluster is now added to the final cluster catalog, 
and we proceed by removing it from the set of filtered maps $J_{\thetac}$ and 
$\L_{\thetac}$ before returning to step 2. To this end, we 
construct beforehand a 2D array (library) of un--normalized, filtered cluster 
templates (postage--stamp maps)
\begin{equation}
{\cal T}_{\thetac,\thetaf}(\vec{x}) = \int d^2x'\; \Psi_{\thetaf}
	(\vec{x}'-\vec{x})\Tc(\vec{x}')
\end{equation}
with the cluster centered in the map.  Note that $\thetac$ runs over core
radius and $\thetaf$ over filter scale. 
At each filter scale $\thetaf$, we place the 
normalized template $\hat{\yo}{\cal T}_{\thetac,\thetaf}$ on the cluster 
position $(x,y)$ and subtract it from the map.  The library of filtered
templates allows us to perform this step rapidly.

\begin{table}[htbp]
\centering
\begin{tabular}{c|c|c|c|c}
  Type & Frequencies & Res. {\it fwhm} & Inst. noise          & Area \\
    \  &    [GHz]    &   [arcmin]      & [$\mu {\rm K/beam}$] & [${\rm deg}^{2}$] \\ \hline \hline
   AMI &      15     &       1.5       &         8            &      10            \\ \hline
   SPT &     150     &       1         &      10              &      \              \\ 
    \  &     220     &       0.7       &      60              &     4000            \\
    \  &     275     &       0.6       &      100             &       \             \\ \hline
   Planck &  143     &       7.1       &       6              &       \             \\
     \    &  217     &       5         &       13             &     41253           \\
     \    &  353     &       5         &       40             &       \
\end{tabular}
\caption[Experiments]
{Characteristics of the three types of experiments considered. 
We run our extraction method on
100 sky patches of $3 \times 3$ square degrees (for AMI and SPT) and 
$12 \times 12$ square degrees (for Planck).  
}
\label{tab:experiments}
\end{table}

We then return to step 2 and repeat the process until there are no 
remaining candidate pixels.  Thus, clusters are added to the catalog 
while being subtracted from the maps one at a time, thereby de-blending
the sources.  By pulling off the brightest clusters first, we aim
to minimize uncertainty in the catalog photometric parameters.  Nevertheless,
it must be emphasized that the entire procedure relies heavily on the
use of templates and that real clusters need not conform to the chosen
profiles. We return to the effects of cluster morphology below.

In the end, we have a cluster catalog with positions $(x,y)$, central
Compton $y$ parameters, sizes $\theta_c$ and fluxes $Y$.

\section{Cluster recovery}

We tested our catalog construction method on simulated observations 
of the three representative types of SZ survey specified in Table~\ref{tab:experiments}.  The simulations include SZ emission, primary CMB
anisotropy and instrumental noise and beam smearing.  We do not include diffuse Galactic foregrounds in this study.  We begin in this section with the
ideal case where the filter perfectly matches the simulated clusters
(spherical $\beta$--model profiles) and in the absence of extragalactic point sources.  We return to the additional effects of cluster morphology and
point source confusion in Section 5.

The simulated maps are generated by Monte Carlo.  We first create
a realization of the linear matter distribution in a large box using
the matter power spectrum.  Clusters are then distributed according
to their expected number density, given by the mass function, and bias
as a function of mass and redshift.  We also give each cluster a 
peculiar velocity consistent with the matter distribution according to linear theory.  The simulations thus featuring cluster spatial and 
velocity correlations accurate first order, which is a reasonable approximation
on cluster scales. In this paper, we use these simulations but we do not study the impact of the correlations on the detection method, leaving this issue to forthcoming work.

The cluster gas is modeled by a spherical isothermal $\beta$--profile with $\beta=2/3$ and $\thetac/\thetav=0.1$, where $\thetav$ is the angular projection
of the virial radius and which varies with cluster mass and redshift following a self-similar relationship. We choose an $M-T$ relation consistent with the local abundance of X--ray clusters and our value of $\sigma_8$, given below (Pierpaoli et al.~\cite{pier04}). 
Finally, we fix the gas mass fraction at $\fgas=0.12$ 
(e.g., \cite{mohr99}). The input catalog consists of clusters with 
total mass $M>10^{14}M_{\odot}$, 
which is sufficient given the experimental characteristics listed in Table~\ref{tab:experiments}.  \cite{del02} describe the simulation method in more detail.

We generate primary CMB anisotropies using the power spectrum calculated
by CMBFAST\footnote{{\tt http://cmbfast.org/}} (\cite{sel96}) for a flat concordance model with $\OmM=0.3=1-\OmL$ (\cite{spe03}), Hubble constant 
$\Ho = 70$~km/s/Mpc (\cite{hstkey}) and a power 
spectrum normalization $\sigma_8=0.98$.  As a last step we smooth the map 
with a Gaussian beam and add Gaussian white noise to model 
instrumental effects\footnote{The 3--year WMAP results, published after the work presented here
was finished, favor a significantly lower value
of $\sigma_8$ (\cite{spe06}).  This could lower the total number of clusters in our simulations by up to a 
factor of $\sim 2$.  As we are interested here in catalog recovery, where we compare output to input
catalogs, this change should only cause relatively minor changes to our final results.}. 

We simulate maps that would be obtained from the proposed surveys 
listed in Table~\ref{tab:experiments}. The first is an 
AMI\footnote{{\tt http://www.mrao.cam.ac.uk/telescopes/ami/index.html}}--like 
experiment (\cite{jon02}), a single frequency, high resolution 
interferometer; the sensitivity corresponds to a one--month integration 
time per 0.1 square degree (\cite{kne01}).  
The SPT\footnote{{\tt http://astro.uchicago.edu/spt/}}--like 
experiment (\cite{ruh04}) is a high resolution, multi--band bolometer array.  
We calculate the noise levels assuming 
an integration time of 1 hour per square degree, and a split of 
2/3, 1/6, 1/6 of the 150, 220, 275~GHz channels for the 1000 detectors 
in the focal plane array (\cite{ruh04}).
Finally, we consider the space--based 
Planck\footnote{{\tt http://www.rssd.esa.int/index.php?project=PLANCK}}--like 
experiment, with a nominal sensitivity for a 
14 month mission.  For the AMI and SPT maps we use pixels\footnote{Pixel 
sizes are at least 2 times smaller than the best channel of each experiment.} of $30 \, {\rm arcsec}$, while for Planck the pixels are $2.5 \, {\rm arcmin}$.

We simulate 100 sky patches of $3\times 3$ square degrees
for both AMI and SPT, and of $12\times 12$ square degrees for Planck.
This is appropriate given the masses of detected clusters in each experiment.
In practice, AMI will cover a few square degrees, similar to the 
simulated patch, while SPT will cover 4000 square degrees and Planck will
observe the entire sky.  Thus, the surveys decrease in sensitivity 
while increasing sky coverage from top to bottom in Table~\ref{tab:counts}
(see also Table~\ref{tab:experiments}).

\begin{figure}[htb]
\includegraphics[scale=0.5]{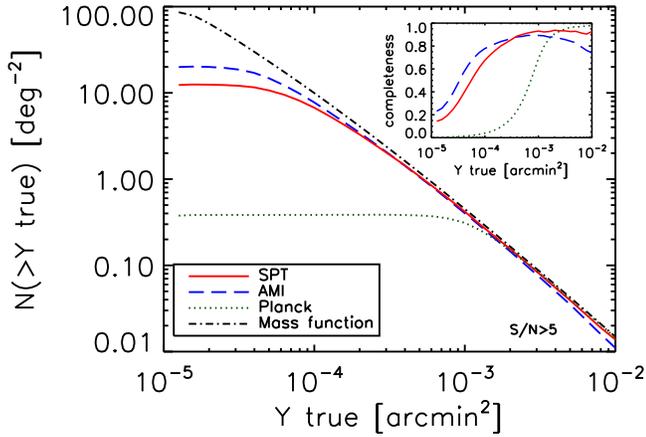}
\caption{Cluster counts $N(>Y)$ per square degree as a function of
true SZ flux $Y$ for a threshold of $S/N>5$. The dash--dotted 
black line gives the cluster counts from the mass function (\cite{jen01}).
The dashed blue line gives the recovered cluster counts  
for AMI, the red solid line for SPT and the dotted green line for Planck.
The inset shows the completeness ratio (relative to the mass function
prediction) for each survey.  All the surveys are significantly incomplete
at their point--source sensitivities (5 times the y--intercept in 
Figure~\ref{fig:cy_tc}).
}
\label{fig:completeness}
\end{figure}

\begin{table}[htbp]
\centering
\begin{tabular}{c|c|c}
 ${\rm deg^{-2}}$ & $S/N>3$ & $S/N>5$ \\ \hline \hline
       AMI        &   44    &    20   \\
        \         &  (38)   &   (16)  \\ \hline
       SPT        &   35    &    12   \\
        \         &  (27)   &   (11)  \\ \hline
      Planck      &  1.00   &   0.38  \\
        \         & (0.84)  &  (0.35)           
\end{tabular}
  \caption[Detected cluster counts]
{Extracted counts/sq. deg. from simulations of the three types of 
survey. The numbers in parenthesis give the counts predicted by our 
analytic cluster model; the difference is due to 
cluster overlap confusion (see text).
}
\label{tab:counts}
\end{table}

\subsection{Association criteria}

An important issue for catalog evaluation is the association between
a detected object (candidate cluster) with a cluster
from the simulation input catalog (real cluster); in other words, 
a candidate corresponds to which, if any, real cluster.  Any 
association method will be imprecise, and estimates of 
catalog completeness, contamination and 
photometric accuracy will unavoidably depend on the choice of
association criteria.  

We proceed as follows: for each detection, we 
look at all input clusters with centers positioned within a distance 
$r={\sqrt 8} \, \times \, d$,  where $d$ is
the pixel size ($d=30 \, \rm{arcsec}$ for AMI and SPT, 
$d=2.5 \, \rm{arcmin}$ for Planck); this covers the neighboring 24 pixels.
If there is no input cluster, then we have a false detection;
otherwise, we identify the candidate with the cluster whose flux is 
closest to that of the detection.  After running through all the 
candidates in this fashion, we may find that different candidates
are associated with the same input cluster. In this case, we only
keep the candidate whose flux is closest to the common input cluster, and 
we flag the other candidates as false detections (multiple detections).

At this stage, some associations may nevertheless be chance alignments. 
We therefore employ a second parameter, $Y_{\rm cut}$:  a candidate
associated with a real cluster of flux $Y<Y_{\rm cut}$ 
is flagged as a false detection. We indicate these false detections
as diamonds in Figures~\ref{fig:photo5_spt}, \ref{fig:photo5_plk}, 
\ref{fig:photo5_ami} and \ref{fig:photo5_true_ami}. The idea is
that such clusters are too faint to have been detected and the 
association is therefore by chance. In the following, we take 
$Y_{\rm cut} = 1.5 \times 10^{-5} \, {\rm
arcmin}^{2}$ for AMI and SPT, respectively, and 
$Y_{\rm cut} = 3 \times 10^{-4} \, {\rm arcmin}^{2}$ for Planck.
Note that these numbers are well below the point--source sensitivity
(at S/N=5) in each case (see below and Figure~\ref{fig:cy_tc}).

\subsection{Completeness}

\begin{figure}[htb]
\includegraphics[scale=0.5]{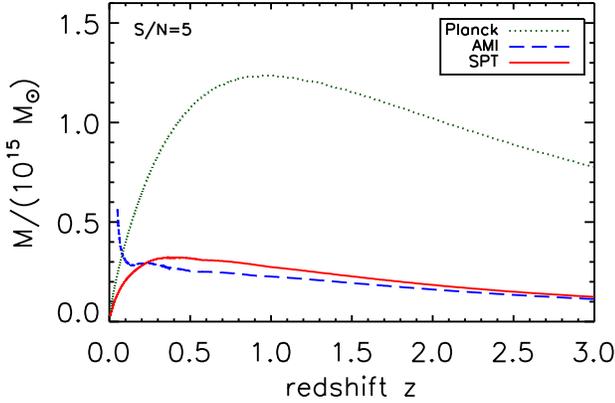}
\caption{Mininum detectable cluster mass as a function of redshift, $M(z)$, corresponding to $S/N=5$ for the three experiments discussed in the text.  The rise at low redshift for the single--frequency (AMI) curve is caused by confusion with primary CMB
anisotropy.
}
\label{fig:mzlim}
\end{figure}

Figure~\ref{fig:completeness} shows completeness for the 
three experiments in terms of {\em true} integrated $Y$, 
while Table~\ref{tab:counts} summarizes the counts.  In Figure~\ref{fig:mzlim}
we give the corresponding limiting mass as a function of redshift.
Given our cluster model,
AMI, SPT and Planck should find, respectively, about 16, 11 and 0.35 clusters/deg.$^2$ at a $S/N>5$; these are the numbers given in parentheses in 
Table~\ref{tab:counts}.  Cluster overlap confusion accounts for the fact 
that the actual counts extracted from the simulated surveys are higher:
some clusters that would not otherwise pass the detection cut 
enter the catalog because the filter adds in flux from close neighbors.  

A detection threshold of $S/N=5$ corresponds to a point--source 
sensitivity of just below $Y=5\times 10^{-5}$~arcmin$^2$ for both 
AMI and SPT, as can be read off the left--hand--side of 
Figure~\ref{fig:cy_tc}.  The surveys approach a high level of
completeness only at $Y>10^{-4}$~arcmin$^2$, however, due to the rise
of the selection cut with core radius seen in Figure~\ref{fig:cy_tc}. 
For these high resolution surveys, point--source sensitivity gives 
a false idea of the survey completeness flux limit. 

At the same signal--to--noise threshold,
Planck is essentially complete above $Y \sim 10^{-3} \, {\rm arcmin^2}$ 
and should detect about 0.4 clusters per square degree.  Since most
clusters are unresolved by Planck, the survey reaches a high completeness 
level near the point--source sensitivity.  We also see this from
the small slope of the Planck selection cut in Figure~\ref{fig:cy_tc}.

We emphasize that the surveys (in particular, the high resolution surveys) are not flux limited for any value of $q$, because increasing $q$ simply 
translates the curve in Figure~\ref{fig:cy_tc} along the $y$ axis. However, one can approach a flux--limited catalog by selecting clusters at $S/N>q$ and then cutting the resulting catalog at \mbox{$Y_o>Y_{\rm limit}\equiv Q\sigma_Y(\theta_c=0.1~{\rm arcmin})$}, where the constant $Q>q$.  As $Q$ increases we tend towards a catalog for which $Y\sim Y_o>Y_{\rm limit}$.  In the case
of SPT with $q=3$, for example, we find that large values of $Q$ ($> 10$)
are required to approach a reasonable flux--limited catalog; this construction, however, throws away a very large number of detected clusters.

Although the AMI (single frequency) and 
SPT (multi-band) survey maps have comparable depth, 
SPT will cover $\sim 4000$~sq. degrees, compared to AMI's $\sim 10$~sq. 
degrees.  Planck will only find the brightest clusters, but with 
full sky coverage.  Predictions for the counts suffer from cluster modeling 
uncertainties, but the comparison between experiments is robust and of 
primary interest here.

\begin{figure}[htb]
\includegraphics[scale=0.5]{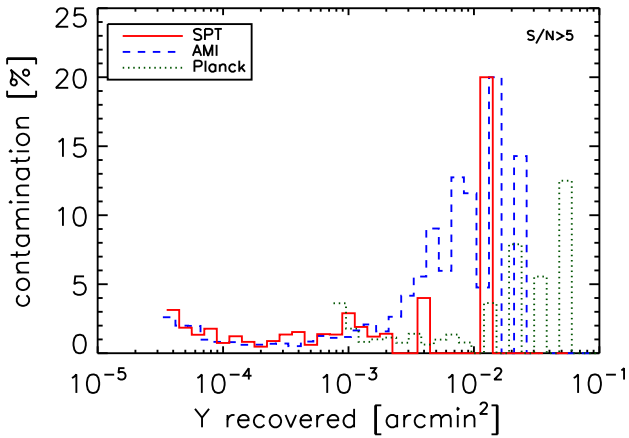}
\caption{Contamination as a function of the core radius $\theta_c$ for 
the three experiments and for ${\rm S/N}>5$.
}
\label{fig:contam}
\end{figure}

\subsection{Contamination}
\begin{figure}[htb]
\includegraphics[scale=0.5]{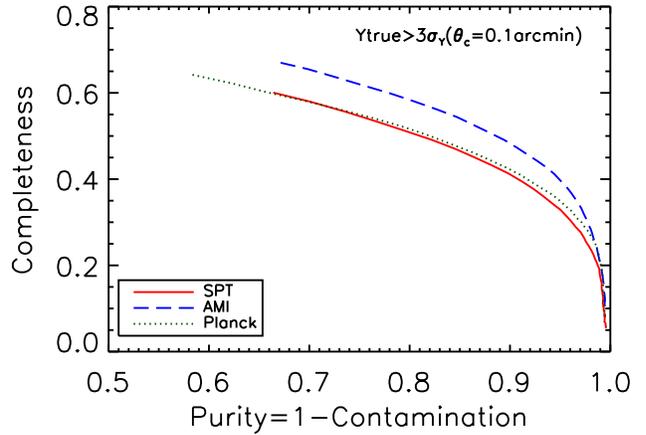}
\caption {Completeness-Purity plot. For each curve, $q$ varies from 3 (top-left) to 10 (bottom-right). For each experiment, the input catalog contains clusters with true flux greater than three times the point source sensitivity ($Y{\rm true}>2.2 \times 10^{-5}$~arcmin$^2$ for AMI,  $Y{\rm true}>2.6 \times 10^{-5}$~arcmin$^2$ for SPT and $Y{\rm true}>4.8 \times 10^{-4}$~arcmin$^2$ for Planck). See text for details.
}
\label{fig:cp_pssens}
\end{figure}

\label{sec:contam}

\begin{figure}[htb]
\includegraphics[scale=0.5]{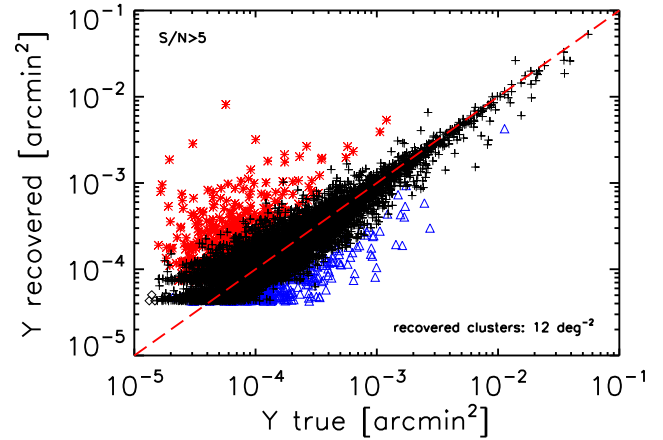}
\caption {Recovered vs. true flux for SPT clusters extracted at ${\rm S/N}>5$
from 100 survey simulations.  The diamonds indicate cluster detections 
with $Y<Y_{cut}$, which we take as false detections. The mean
trend $\Yo(Y)$ has a slight bias (see text) and a roughly constant scatter 
of $\sigma_{log{\Yo}}=0.17$ over the interval in true $Y$ from $10^{-4}$~arcmin$^2$ to 
$4\times 10^{-3}$~arcmin$^2$. The clusters which have their core radii overestimated by a factor of 2 are plotted as red crosses and the clusters
which have their core radii underestimated by a factor of 2 are plotted as blue triangles.
 }
\label{fig:photo5_spt}
\end{figure}

Figure~\ref{fig:contam} shows the contamination level at $S/N>5$ for 
each survey type as a function of {\em recovered} flux $\Yo$.  
The multiband experiments (SPT and Planck) benefit from low 
contamination at all fluxes.  
Single frequency surveys (e.g., AMI), on the other hand, experience a 
slightly higher contamination level at large flux due to confusion from
primary CMB anisotropy.  This confusion also degrades the photometry, 
as we discuss below.  

At $S/N>5$, the AMI, SPT and Planck catalogs have less than 2\% total contamination
rate. These numbers 
increase to $\sim 23$, 20 and $27$ percent, respectively, for AMI, SPT and
Planck at a detection threshold of $S/N>3$.  Note that the total 
contamination rate is an average over the histogram of Figure~\ref{fig:contam}
weighted by the number of objects in each bin; thus, the higher contamination
at large flux is down--weighted in the total rate. 

In all cases, the
contamination rate is higher than expected from pure Gaussian noise 
fluctuations; there is an important contribution from cluster--cluster 
confusion (residuals from cluster subtraction and overlaps).  
We expect even higher contamination rates in practice, because of 
variations in cluster morphology around the filter templates.   
We quantify this latter effect below.

A useful summary of these results is a completeness--purity plot, as 
shown in Figure~\ref{fig:cp_pssens}. Proper comparison of the 
different experiments requires an appropriate choice of input catalog 
used to define the completeness in this plot.  Here, we 
take the input catalog as all clusters with (true) flux geater than 
three times the point source sensitivity for each experiment. 
If the clusters were point sources and the detection method perfect (i.e. not affected by confusion), the completeness would be 1 for $q=3$ in the top-left corner. These curves summarize the efficiency of our cluster detection method; however, they give no information on the photometric capabilities of the experiments.

\subsection{Photometry}
\label{sec:photo}

We now turn to the important, but often neglected issue of cluster 
SZ photometry.  The ability of a SZ survey to constrain cosmology 
relies on application of the $Y-M$ relation.  As mentioned, we expect 
the {\em intrinsic} (or {\em true}) flux to tightly correlate with cluster 
mass (\cite{bar01}), as indeed borne out by numerical simulations 
(\cite{dasilva_etal04, motl_etal05, nagai05}).  Nevertheless, unknown
cluster physics could affect the exact form and normalization of 
the relation, pointing up the necessity of an empirical calibration (referred
to as survey {\em calibration}), either with the survey data itself 
(self--calibration, \cite{hu03,maj_mohr04,lima_hu04,lima_hu05}) or using 
external data, such as lensing mass estimates (\cite{bartel01}) (although  
the latter will be limited to relatively low redshifts).

Photometric measurement
accuracy and precision is as important as cluster physics in this context: 
what matters in practice is the relation between {\em recovered} SZ flux $\Yo$
and cluster mass $M$.  Biased SZ photometry (bias in the $Y-\Yo$) relation
will change the form and normalization of the $\Yo-M$ relation and noise will
increase the scatter.  One potentially important source of photometric 
error for the matched filter comes from cluster morphology, i.e., 
the fact that cluster profiles do not exactly follow the filter shape 
(see Section 5).

Survey calibration will help remove the bias, but
with an ease that depends on the photometric scatter: large scatter will
increase calibration uncertainty and/or necessitate a larger amount 
of external data. In addition, scatter will degrade the final cosmological 
constraints (e.g., \cite{lima_hu05}).  
Photometry should therefore be considered an important 
evaluation criteria for cluster catalog extraction methods. 

\begin{figure}[htb]
\includegraphics[scale=0.5]{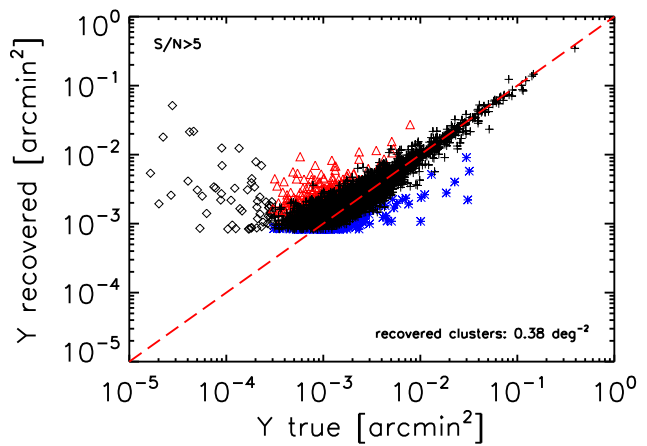}
\caption{Recovered vs. true flux for Planck clusters extracted at 
${\rm S/N}>5$ from 100 survey simulations.  The diamonds indicate 
cluster detections with $Y<Y_{cut}$, which we take as false detections.  
The mean trend $\Yo(Y)$ has a slight bias (see text) and a 
roughly constant scatter of $\sigma_{log{\Yo}}=0.13$ over the interval in true $Y$ 
from $2\times 10^{-3}$~arcmin$^2$ to $2\times 10^{-2}$~arcmin$^2$.The clusters which have their core radii overestimated by a factor of 2 are plotted as red crosses and the clusters
which have their core radii underestimated by a factor of 2 are plotted as blue triangles.
}
\label{fig:photo5_plk}
\end{figure}

Consider, first, SPT photometry.  Figure~\ref{fig:photo5_spt} shows the
relation between observed (or recovered) flux $\Yo$ and true flux $Y$ 
for a detection threshold of $S/N>5$.  Fitting for the average trend
of $\Yo$ as a function of $Y$, we obtain
\begin{displaymath}
{\rm log}\Yo=0.96 {\rm log}Y-0.15
\end{displaymath} 
over the interval $10^{-4}$~arcmin$^2$~$< Y < 4\times 10^{-3}$~arcmin$^2$,
with $\Yo$ and $Y$ measured in arcmin$^2$.
There is a slight bias in that the fit deviates somewhat from the 
equality line, but the effect is minor.  
Below this flux interval, the fit curls upward in a form 
of Malmquist bias caused by the $S/N$ cut (seen as the sharp lower edge
on $\Yo$).  The lack of any significant bias is understandable in this ideal case where the filter perfectly matches the cluster SZ profile.
Cluster morphology, by which we mean a mismatch between the cluster SZ profile and the matched filter template), can induce bias; we return to this issue
in Section 5.

The scatter about the fit is consistent with a Gaussian
distribution with a roughly constant 
standard deviation of $\sigma_{log{\Yo}}=0.17$ over the entire interval.

The scatter is a factor of 10 larger than expected from instrumental noise
alone, which is given by the selection curve in Figure~\ref{fig:cy_tc}.
Uncertainty in the recovered cluster position, core radius and 
effects from cluster--cluster confusion all strongly influence the 
scatter.  Photometry precision, therefore, cannot be predicted from 
instrumental noise properties alone, but only with simulations 
accounting for these other, more important effects.  

Figure~\ref{fig:photo5_plk} shows the photometry for the Planck survey.
Apart from some catastrophic cases (the diamonds), the photometry is
good and fit by
\begin{displaymath}
{\rm log}\Yo = 0.98 {\rm log}Y - 0.07 
\end{displaymath}
over the interval $2\times 10^{-3}$~arcmin$^2 < Y < 2\times 10^{-2}$~arcmin$^2$
($\Yo$, $Y$ measured in arcmin$^2$).  The dispersion is 
$\sigma_{log\Yo}=0.13$, roughly constant over the same interval.  For 
unresolved clusters, this scatter is $\sim 5$ times larger than the 
expected instrumental--induced scatter. 
The brightest diamonds in the Figure correspond to real clusters 
with positional error larger than the association criteria $r$.  
As a consequence, the candidates are falsely associated with a small,
nearby cluster, unrelated to the actual detected object.

We emphasize that the observational scatter in the $\Yo-Y$ relation 
for both SPT and Planck dominates the intrinsic scatter of less 
than 5\% seen in the $Y-M$ relation from numerical simulations 
(\cite{dasilva_etal04, motl_etal05}).  

\begin{figure}[htb]
\includegraphics[scale=0.5]{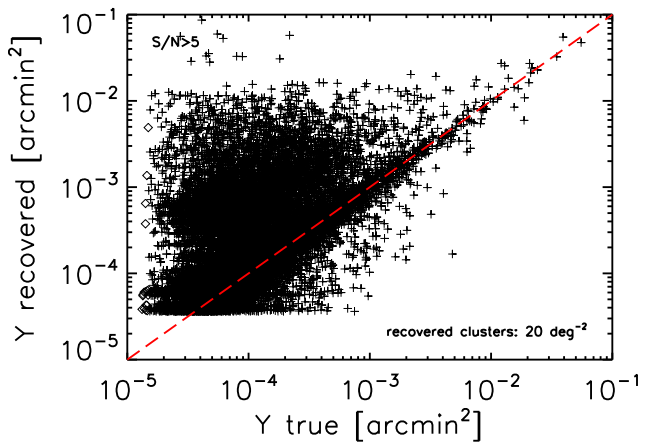}
\caption {Recovered vs. true flux for AMI clusters extracted at ${\rm S/N}>5$
from 100 survey simulations.  The diamonds indicate cluster detections 
with $Y<Y_{cut}$, which we take as false detections.  The extremely  
large dispersion in recovered flux results from a bimodal distribution 
caused by an inability to determine the core radius of detected clusters.
This inability is due to confusion from primary CMB anisotropy, as 
demonstrated in Figure~\ref{fig:photo5_ami_histo}.
Figure~\ref{fig:photo5_true_ami} shows that reasonable photometry is
possible if the core radius can be accurately determined. 
This problem is specific to single--frequency surveys that are unable to
spectrally remove primary CMB anisotropy.
}
\label{fig:photo5_ami}
\end{figure}

\begin{figure}[htb]
\includegraphics[scale=0.5]{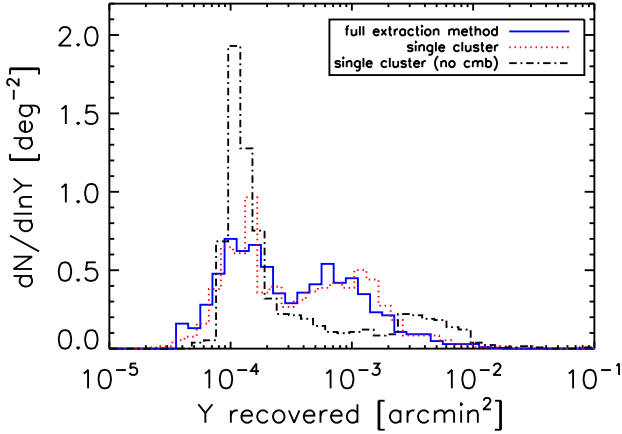}
\caption{The full blue histogram gives the cluster counts from 
figure~\ref{fig:photo5_ami} in the bin 
($10^{-4}<Y<2.10^{-4}$, $0.25<\theta_c<0.35$). We have added the 
cluster counts obtained from the size and flux estimation of a 
single cluster ($Y=1.5 \times 10^{-4}$, $\theta_c=0.3$)
at a known position through 1000 simulations. SZ cluster background 
maps and the instrumental beam and noise are included.
Two cases are considered : with primary CMB (dotted red histogram) 
and without primary CMB (dash--dotted black line). The double bump
in Y recovery is visible when the primary CMB is present and disappears 
when it's removed showing that the primary CMB power spectrum
is the cause of the double bump.
}
\label{fig:photo5_ami_histo}
\end{figure}

\begin{figure}[htb]
\includegraphics[scale=0.5]{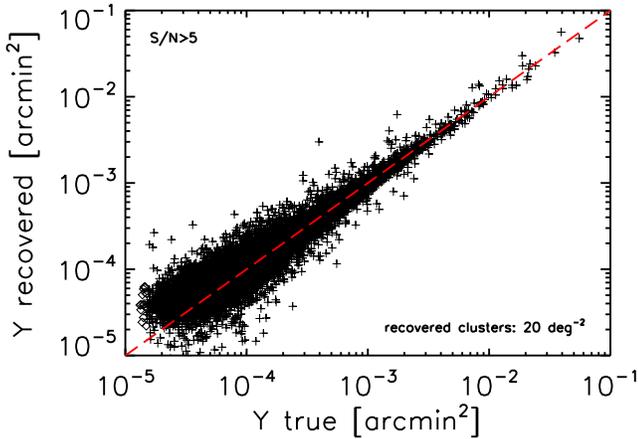}
\caption {Single--frequency photometry when we artificially set the
core radii of detected clusters to their true values from the input 
catalog. The dispersion decreases dramatically, demonstrating that
the inability to recover the core radius is the origin of the 
bad photometry seen in Figure~\ref{fig:photo5_ami}.
}
\label{fig:photo5_true_ami}
\end{figure}

We now turn to single frequency surveys, which Figure~\ref{fig:photo5_ami} 
shows to have seriously compromised photometry. The distribution
at a given true flux $Y$ is in fact bimodal, as illustrated by the solid
blue histogram in Figure~\ref{fig:photo5_ami_histo} that
gives the distribution of the recovered flux $\Yo$ for clusters with 
true flux and core radius in a bin centered on 
$Y=1.5\times 10^{-4}$~arcmin$^2$ and $\thetac=0.3$~arcmin. 
We have traced this  effect to an inability to accurately determine the 
core radius of the candidate clusters.  We demonstrate this in 
Figure~\ref{fig:photo5_true_ami} by artificially setting the candidate
core radius to its true value taken from the associated input
cluster; the photometry now cleanly scatters about the mean trend.

This inability to determine the core radius mainly arises from confusion with
primary CMB anisotropy, as we now show using 
Figure~\ref{fig:photo5_ami_histo}.   
We performed 1000 simulations of a single cluster 
($Y=1.5\times 10^{-4}$~arcmin$^2$, $\thetac=0.3$~arcmin) placed at
the middle of a beam--convolved map containing background SZ clusters 
(from our general simulations), primary CMB anisotropy and 
instrumental noise.  We then estimate its core radius and flux with 
our matched filters centered on the known position (to avoid any
positional uncertainty) and trace the histogram of resulting
measured flux.  This is the red dot--dashed histogram in the figure, 
which displays a bi--modality similar to that of the blue solid 
histogram.  We then follow the same procedure after first removing the
primary CMB anisotropy from the simulated map.  
The resulting histogram of recovered flux is given
by the black dot--dashed line with much less pronounced bimodality.
The remaining tail reaching towards high flux is caused by cluster--cluster
confusion.

With their additional spectral information, multiband surveys remove
the primary CMB signal, thereby avoiding this source of confusion.
{\em The result suggests that follow--up observations
of detected clusters at a second frequency will be required for 
proper photometry}; without such follow--up, the scientific power
of a single frequency survey may be seriously compromised, as can
be appreciated from inspection of Figure~\ref{fig:photo5_ami}.

\section{Additional Effects}

As emphasized, our previous results follow for a filter that perfectly 
matches the (spherical) clusters in our simulations and in the
absence of any point sources.  In this section we examine 
the effects of both cluster morphology and point sources.  

We find that cluster morphology has little effect on catalog completeness, 
but that it does increase contamination.  More importantly, it can bias 
photometric recovery, although it does not significantly increase the scatter.  This bias changes the observed $Y-M$ relation from its intrinsic form, adding
to the modeling uncertainty already caused by cluster gas physics. 
For this reason, the relation must be calibrated in order to use 
the SZ catalog for any cosmological study.  The observational bias would be removed during this calibration step. 

Completeness is the most affected by point source confusion, decreasing somewhat 
for the multi--band surveys in the presence of IR point
sources.  The level of confusion for the single frequency survey remains
highly uncertain due to the unknown point source counts at low flux densities.
Contamination and photometry are essentially unaffected.

\subsection{Cluster Morphology}

\begin{figure}[htb]
\includegraphics[scale=0.5]{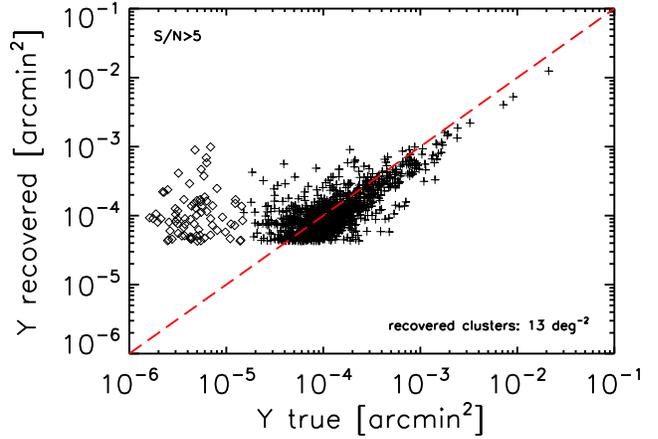}
\caption {Photometry for the SPT catalog from the N--body simulations. 
Cluster morphology (mismatch between the filter profile and the 
actual cluster SZ profile) clearly induces a bias between the recovered
and true SZ flux.  The scatter, on the other hand, is not very 
affected, as can be seen in comparing with Figure~\ref{fig:photo5_spt}.
}
\label{fig:spt_photo_nbody}
\end{figure}

To assess the influence of cluster morphology, we ran our 
catalog extraction algorithm on maps constructed from numerical simulations.
We use the simulations presented by Schulz \& White (\cite{sch03}) and kindly
provided to us by M. White. Their simulations follow dark matter clustering with
a N--body code in a flat concordance cosmology, 
and model cluster gas physics with semi--analytical 
techniques by distributing an isothermal gas of mass fraction $\OmB/\OmM$ 
according to the halo dark matter distribution. For details, see Schulz \& White.
In the following, we refer to these simulations as the ``N--body'' simulations.

We proceed by comparing catalogs 
extracted from the N--body map to those from a corresponding 
simulation made with spherical clusters.  The latter is constructed by applying
our spherical $\beta$--model gas distribution to the cluster halos taken 
from the N--body simulation and using them as input to our Monte Carlo
sky maps.  In the process, we renormalize our $Y-M$ relation to the one 
used in the N--body SZ maps.  We thus obtain two SZ maps containing the same
cluster halos, one with spherical clusters (referred to hereafter as the ``$\beta$--model''
maps) and the other with more complex cluster morphology (the N-body maps).  
Comparison of the catalogs extracted from the two different types of
simulated map gives us an indication of the sensitivity of
our method to cluster morphology.  We make this comparative study only for the
SPT and Planck like surveys.

\begin{figure}[htb]
\includegraphics[scale=0.5]{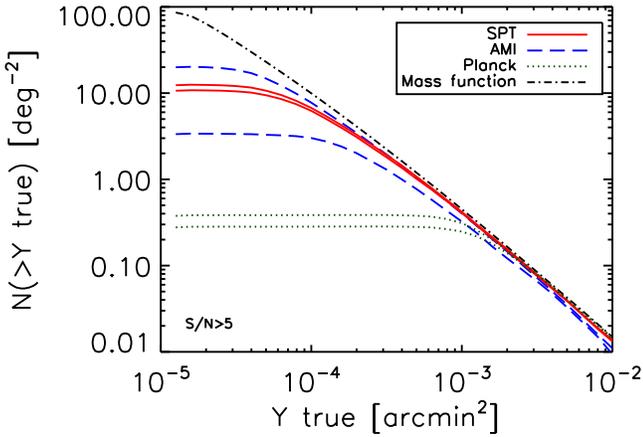}
\caption {Integrated cluster counts for the three types of survey.  The upper curve
in each pair reproduces the results of Figure~\ref{fig:completeness}, while the
lower curve shows the effect of point source confusion.  Despite the large 
IR point source population, multiband surveys efficiently eliminate confusion.
The AMI--like survey is, on the other hand, strongly affected.  This latter effect
remains uncertain due to a lack of information on the
faint end of the radio point source counts (see text).
}
\label{fig:completeness_sources}
\end{figure}

Catalog completeness is essentially unaffected by cluster morphology; the integrated counts, for example, follow the same curves shown in 
Figure~\ref{fig:completeness} with very little deviation, the only 
difference being a very small decrease in the Planck
counts at the lowest fluxes.  The effect, for example, is smaller than 
that displayed in Figure~\ref{fig:completeness_sources} due to point 
source confusion (and discussed below).

Non--trivial cluster morphology, however, does significantly 
increase the catalog contamination rate; for example, in the SPT survey 
the global contamination rises from less than 2\% to 13\% at $S/N=5$ for the N--body simulations.  We trace this to residual flux left in the maps after 
cluster extraction: cluster SZ signal that deviates from the assumed spherical $\beta$--model filter profile remains in the map and is 
picked up later as new cluster candidates.  Masking those 
regions where a cluster has been previously extracted (i.e.,
forbidding any cluster detection) drops the contamination to 4\% (SPT case), but causes a decrease 
of 2.8 clusters per square degree in the recovered counts; this technique would also 
have important consequences for clustering studies.

From Figure~\ref{fig:spt_photo_nbody}, we clearly see that cluster
morphology induces a bias in the photometry.  This arises from the fact
that the actual cluster SZ profiles differ from the template adopted
for the filter.  The differences are of two types: an overall difference 
in the form of radial profile and local deviations about the average radial profile due to cluster substructure.  It is the former that is primarily
responsible for the bias.  In our case, the N--body simulations have 
much more centrally peaked SZ emission than the filter templates, which
causes the filter to systematically underestimate the total SZ flux.
Cluster substructure will increase the scatter about the mean 
$\Yo-Y$ relation.  This latter effect is not large, at least for 
the N--body simulations used here, as can be seen by comparing the scatter
in Figures~\ref{fig:spt_photo_nbody} and \ref{fig:photo5_spt}.  

We emphasize, however, that the quantitative effects on photometry
depend on the intrinsic cluster profile, and hence are subject to
modeling uncertainty.  The simulations used here do not include gas physics and simply assume that the gas follows the dark matter.  The real 
bias will depend on unknown cluster physics, thus adding to the
modeling uncertainty in the $Y-M$ relation.  This uncertainty, due to
both cluster physics and the photometric uncertainty discussed here, 
must be dealt with by empirically calibrating the relation, either with 
external data (lensing) and/or internally (self--calibration).  
  
\subsection{Point Sources}

We next examine the effect of point sources.  In a previous paper 
(\cite{bar05}, hereafter BM) we studied their influence on survey detection
sensitivity.  We extend this work to our present study in this section.  

Low frequency surveys, such as our AMI example, contend with an important radio source population, while higher frequency bolometer surveys face a large population of IR sources.  Radio source counts down to the sub--mJy flux levels relevant for SZ surveys are unfortunately poorly known.  The IR counts are somewhat better constrained at fluxes dominating the fluctuations in the IR background, although at higher frequencies (850~microns) than those used in SZ surveys; an uncertain extrapolation in frequency is thus necessary.  

For the present study, we use the radio counts fit by \cite{knox04} to a combination of 
data from CBI, DASI, VSA and WMAP (see also Eq.~6 in BM), and
IR counts fit to blank--field SCUBA observations at 850~microns by \cite{borys03}
(and given by Eq.~8 in BM).  We further
assume that all radio sources brighter than 100~$\mu$Jy have been subtracted from 
our maps at 15~GHz (AMI case); this is the target sensitivity of the long baseline
Ryle Telescope observations that will perform the source subtraction for AMI.
No such explicit point source subtraction is readily available for the higher
frequency bolometer surveys; they must rely solely on their frequency coverage to
reduce point source confusion.  We therefore include all IR sources in our 
simulations, and fix their effective spectral index $\alpha=3$ with no 
dispersion\footnote{As discussed in BM, any dispersion has only a small effect 
on survey sensitivity}. 
We refer the reader to BM for details of our point source model. 
Note that for this study we use the spherical cluster model for direct comparison 
to our fiducial results.

Figure~\ref{fig:completeness_sources} compares the integrated counts from
Figure~\ref{fig:completeness} (upper curve in each case) to those extracted from 
the simulations including point sources (lower curves).  We see that point source confusion only slightly decreases the completeness of the multiband surveys, but greatly affects the single frequency survey.   

In the case of SPT, this is because point source
confusion remains modest compared to the noise: the two are comparable
at 150~GHz, but the noise power rises more quickly with frequency than
the confusion power (see BM for details) -- in other words, the noise
is bluer than the confusion.  This is an important consideration when
looking for the optimal allocation of detectors to the observation bands.  

For Planck, confusion power dominates at all frequencies, but the spectral 
coverage provides sufficient leverage to control it.  In this light, 
it must be emphasized that we only include three astrophysical signals 
(SZ, CMB \& point sources) in these simulations, so that three observation
bands are sufficient.  In reality, one will have to deal with other
foregrounds, e.g., diffuse Galactic emission, which will require the use
of additional observation bands.

The single frequency observations, on the other hand, are strongly affected.  
This is consistent with the estimate in BM (Eq.~15) placing confusion noise 
well above instrumental noise for the chosen point source model and source subtraction threshold.  We emphasize the uncertainty in this estimate, 
however: in BM we showed, for example, that a model with flattening counts 
has much lower source confusion while remaining consistent with the 
observed counts at high flux densities.  The actual confusion level remains to be
determined from deeper counts at CMB frequencies 
(see \cite{wald_etal03,wald_pooley04} for recent deep counts at 15~GHz).

Contamination in the multiband surveys is practically unaffected by point
source confusion. For AMI we actually find a lower contamination rate, an 
apparent gain explained by the fact that the catalog now contains only the 
brighter SZ sources, due to the lowered sensitivity caused by point source
confusion.

The photometry of the multiband surveys also shows little effect from
the point sources.  Fits to the recovered flux vs. true flux relation do not
differ significantly from the no--source case, and the dispersion remains essentially the same.  This is consistent with the idea that point source
confusion is either modest compared to the noise (SPT) or controlled
by multiband observations (Planck).


\section{Discussion and Conclusion}

We have described a simple, rapid method based on matched multi--frequency filters
for extracting cluster catalogs from SZ surveys. We assessed its performance when 
applied to the three kinds of survey listed in Table~\ref{tab:experiments}.
The rapidity of the method allows us to run many simulations of each survey
to accurately quantify selection effects and observational uncertainties.  
We specifically examined catalog completeness, contamination rate and photometric 
precision.

Figure~\ref{fig:cy_tc} shows the cluster selection criteria in terms of total
SZ flux and source size.  It clearly demonstrates that SZ surveys, in particular
high resolution ground--bases surveys, will not be purely flux limited, something
which must be correctly accounted for when interpreting catalog statistics
(Melin et al.~\cite{mel05}).   

Figure~\ref{fig:completeness} and Table~\ref{tab:counts} 
summarize the expected yield for each survey.  The counts roll off at 
the faint end well before the point--source flux limit (intercept of the curves
in Figure~\ref{fig:cy_tc} multiplied by the S/N limit) 
even at the high detection threshold of S/N=5; 
the surveys loose completeness precisely because they are not purely flux--limited.
These yields depend on the underlying 
cluster model and are hence subject to non--negligible uncertainty.  
They are nonetheless indicative, and in this work we focus on the nature of 
observational selection effects for which the exact yields are of secondary 
importance.   

At our fiducial S/N=5 detection threshold, overall catalog contamination remains 
below 5\%, with some dependence on SZ flux for the single frequency survey
(see Figure~\ref{fig:contam}).  The overall contamination rises to between
20\% and 30\% at S/N$>$3.  We note that the contamination rate is always
larger than expected from pure instrumental noise, pointing to the influence
of astrophysical confusion.

We pay particular attention to photometric precision, an issue often neglected
in discussions of the scientific potential of SZ surveys. Scatter plots for the
recovered flux for each survey type are given in Figures~\ref{fig:photo5_spt},
\ref{fig:photo5_plk} and \ref{fig:photo5_ami}.  In the two multiband surveys, the
recovered SZ flux is slightly biased, due to the flux cut, with a dispersion
of $\sigma_{log\Yo}=0.17$ and $\sigma_{log\Yo}=0.13$ for SPT and Planck, respectively.  
This observational dispersion
is significantly larger than the intrinsic dispersion in the $Y-M$ relation predicted 
by hydrodynamical simulations.  This uncertainty must be properly accounted for
in scientific interpretation of SZ catalogs; specifically, it will degrade survey 
calibration and cosmological constraints. 

Even more importantly, we found that astrophysical confusion seriously compromises 
the photometry of the single frequency survey (Figure~\ref{fig:photo5_ami}).  
The histogram in Figure~\ref{fig:photo5_ami_histo} shows that the recovered flux 
has in fact a bimodal distribution.  We traced the effect to an inability 
to determine source core radii in the presence of primary CMB anisotropy.
If cluster core radius could be accurately measured, e.g., with X--ray follow--up,
then we would obtain photometric precision comparable to the multiband surveys
(see Figure~\ref{fig:photo5_true_ami}).  This confusion can also be removed 
by follow--up of detected sources at a second radio frequency (e.g., 90~GHz).
Photometric uncertainty will therefore be key limiting factor in single frequency
SZ surveys.

All these results apply to the ideal case where the filter exactly matches the
(simulated) cluster profiles.  We then examined the potential impact of cluster morphology and point sources on these conclusions.  

Using N--body simulations, we found that cluster morphology has little effect on catalog completeness, but that it does increase the contamination rate and bias the photometry.  The increased 
contamination is caused by deviations from a smooth radial SZ profile that
appear as residual flux in the maps after source extraction.  More importantly,
the photometry is biased by the mismatch between the filter template and
the actual cluster profile.  This observational bias adds to the modeling uncertainty in the $Y-M$ relation, which will have to be empirically 
determined in order to use the catalog for cosmology studies.

As shown by Figure~\ref{fig:completeness_sources}, 
point sources decrease survey completeness.  The multiband surveys effectively 
reduce IR point source confusion and suffer only a small decrease.  Radio source
confusion, on the other hand, greatly decreased the completeness of the single
frequency survey.  This is consistent with the expectation that, for our adopted
radio point source model and source subtraction threshold, point source 
confusion dominates instrumental noise.  Modeling uncertainty here is, however,
very large: radio source counts are not constrained at relevant fluxes 
($\sim 100\;\mu$Jy), which requires us to extrapolate counts from mJy levels
(see BM for a more detailed discussion). 

Surveys based on the SZ effect will open a new window onto the high redshift 
universe.  They inherit their strong scientific potential from the unique characteristics of
the SZ signal.  Full realization of this potential, however, requires understanding
of observational selection effects and uncertainties.  Overall, multiband surveys
appear robust in this light, while single frequency surveys will most likely
require additional observational effort, e.g., follow--up in other wavebands,
to overcome large photometric errors caused by astrophysical confusion with primary
CMB anisotropy.

\begin{acknowledgements}
We thank T. Crawford for useful comments on matched filters and information about SPT, and A. Schulz and M. White for kindly providing us with their N-body simulations. We are also grateful to the anonymous referee for helpful and insightful comments. JBM wishes to thank L. Knox, the Berkeley Astrophysics group and E. Pierpaoli for discussions on the detection method, and D. Herranz and the Santander group for discussions on matched filters.  JBM was supported at UC Davis by the National Science Foundation under Grant No. 0307961 and NASA under Grant No. NAG5--11098.
\end{acknowledgements}

\end{document}